\documentclass[10pt]{revtex4} 
 
	\usepackage{color} 

	\usepackage{graphicx} 
	\usepackage{epsfig}
	\usepackage{epstopdf}

	\usepackage{amsbsy}
	\usepackage{amsfonts}
	\usepackage{amssymb}
\usepackage{amsmath} 
	\usepackage{bm} 

	\newcommand{\tr}[1]{\textrm{tr} \left[ {#1} \right]} 
	\newcommand{\e}[1]{e^{ {#1}}} 

	\newcommand{\be}{\begin{equation}}
	\newcommand{\ee}{\end{equation}}

	\newcommand{\ket}[1]{{\left\vert {#1} \right\rangle}}	
	\newcommand{\bra}[1]{{\left\langle {#1} \right\vert}}	

\begin{document}

\title{Correlations in Quantum Physics}
\author{Ross Dorner}
\email[To whom correspondence should be addressed. E-mail: ]{ross.dorner09@imperial.ac.uk}
\affiliation{Blackett Laboratory, Department of Physics, Imperial College London, London SW7 2AZ, U.K.}
\affiliation{Clarendon Laboratory, Department of Physics, University of Oxford, Parks Road, Oxford OX1 3PU, U.K.}
\author{Vlatko Vedral}
\affiliation{Clarendon Laboratory, Department of Physics, University of Oxford, Parks Road, Oxford OX1 3PU, U.K.,}
\affiliation{Centre for Quantum Technologies, National University of Singapore, 3 Science Drive 2, 117543, Singapore}
\affiliation{Department of Physics, National University of Singapore, 2 Science Drive 3, 117542, Singapore}

\begin{abstract}
We provide an historical perspective of how the notion of correlations has evolved within quantum physics. We begin by reviewing Shannon's information theory and its first application in quantum physics, due to Everett, in explaining the information conveyed during a quantum measurement. This naturally leads us to Lindblad's information theoretic analysis of quantum measurements and his emphasis of the difference between the classical and quantum mutual information. After briefly summarising the quantification of entanglement using these and related ideas, we arrive at the concept of quantum discord that naturally captures the boundary between entanglement and classical correlations. Finally we discuss possible links between discord and the generation of correlations in thermodynamic transformations of coupled harmonic oscillators.
\end{abstract}

\maketitle

\section{Introduction}

Correlations play a prominent role in many-body physics, statistical physics and information technology. In this article, however, we focus on another motivation for studying correlations, namely that they offer the best way of understanding the key differences differences between quantum and classical physics. Quantum mechanics allows for the full range of correlations between events. There is a maximal correlation between earlier and later states of quantum systems undergoing a unitary evolution. Also, if a measurement is repeated successively on the same system, it will always give the same outcome. At the other extreme lie completely uncorrelated quantum events, such as measuring a spin half particle successively in two complementary basis. All other partially correlated events between maximal and zero correlation can also be found in quantum physics. Firstly, we will explain how information theory can be used to quantify correlations in general. We then show how this was applied by Everett and Lindblad to elucidate the quantum measurement process. This is followed by a brief summary of the two main types of quantum correlation, 
 entanglement and quantum discord. Finally, the role of correlations is explored in the field of quantum thermodynamics, where we explore links between correlations and the amount of work done by two couple harmonic oscillators. 

\section{Shannon's information theory}
\label{sec:Shannon}

Given two random variables, $X$ with set of possible outcomes $\{x_i\}$ and $Y$ with the set of possible outcomes $\{y_j\}$, we define the joint probability distribution $P(X=x_i,Y=y_j)\equiv p(x_i,y_j)$ with corresponding marginal probability distributions $P(A=a_i)\equiv p(x_i)=\sum_j p(x_i,y_j)$ and $P(Y=y_j)\equiv p(y_j)=\sum_i p(x_i,y_j)$.
The degree of correlation between the random variables $X$ and $Y$ is encoded within their joint probability distribution and best quantified within the framework of Shannon's information theory~\cite{Shannon}. To that end, we begin our discussion by introducing the Shannon entropy of the random variable $X$, defined as
\begin{align}
S(X) \equiv S(p(x))=  - \sum_i p(x_i) \ln p(x_i),
\label{eq:firstshannonentropy}
\end{align}
and analogously $S(Y)= - \sum_j p(y_j) \ln p(y_j)$. As is well known, the Shannon entropy provides a measure of the uncertainty in a given random variable. By straightforward generalisation of Eq.~\eqref{eq:firstshannonentropy} the uncertainty in the joint distribution of $X$ and $Y$ is defined by 
\begin{align}
S(X,Y) \equiv S(p(x,y)) = -\sum_{i,j} p(x_i,y_j) \ln p(x_i,y_j).
\nonumber
\end{align}
From this, we are able define the the Shannon mutual information,
\begin{align}
I_S(X:Y) = S(p(x))+S(p(y))-S(p(x,y)),
\label{eq:shanmut}
\end{align}
This quantity, being positive, provides us with an intuitive interpretation of correlation between $X$ and $Y$. Namely, that the sum of the uncertainties in $p(x)$ and $p(y)$ alone is greater than the uncertainty in the their joint distribution $p(x,y)$. 
Next, by introducing the Shannon relative entropy,
\begin{align}
S(X||Y)\equiv S(p(x)||p(y))=\sum_{i,j} p(x_i) \ln \frac{p(x_i)}{p(y_j)},
\nonumber
\end{align}
the Shannon mutual information Eq.~\eqref{eq:shanmut} can be rewritten in the equivalent form
\begin{align}
I_S(A:B) = S\left(p(x,y)||p(x)\times p(y)\right).
\nonumber
\end{align}
In this sense, the mutual information
represents a distance between the joint probability distribution ${p(x,y)}$ and the product of
its marginals ${p(x)}\times{p(y)}$. As such, it is intuitively clear that $I_S$ is a good measure of correlations, since it shows how far a joint
distribution is from the corresponding product distribution in which all the correlations have been
destroyed.
Let us now view this from another perspective. Suppose that we wish to know the
probability of observing the event $y_j$ given that the event $x_i$ has been observed. The relevant quantity is the conditional probability, given by
\begin{align}
P(Y =y_j|X=x_i)\equiv p_{x_i}(y_j)=\frac{p(x_i,y_j)}{p(x_i)}.
\nonumber
\end{align}
This motivates us to introduce the conditional entropy, $S_X(Y)$, as
\begin{align}
S_X(Y) &= -\sum_i p(x_i) \sum_j p_{x_i}(y_j) \ln p_{x_i}(y_j)
\nonumber \\
& = -\sum_{i,j} p(x_i,y_j) \ln p_{x_i}(y_j).
\nonumber
\end{align}
The conditional entropy tells us how uncertain we are about the value of $Y$ once we
have learned the value of $X$. Now the Shannon mutual information Eq.~\eqref{eq:shanmut} can
be rewritten once more as
\begin{align}
I_S(X:Y) = S(Y) - S_X(Y) = S(X) - S_Y(X).
\label{clascap}
\end{align}
Thus, as its name indicates, the Shannon mutual information
measures the quantity of information conveyed about the random
variable $X$ ($Y$) through measurements of the random variable $Y$
($X$). As this quantity is positive, it tells us that the initial
uncertainty in $Y$ ($X$) can in no way be increased by making
observations on $X$ ($Y$). Note also that, unlike the Shannon
relative entropy, the Shannon mutual information is symmetric i.e., $I_S(X:Y)=I_S(Y:X)$.

Shannon's original motivation for developing this framework was the practical need of quantifying the channel capacity of classical communication channels. In this scenario, $X$ and $Y$ represent the
message sent down and received from the channel respectively.
Given that process of performing a measurement can be viewed as a communication channel between 
the states of the system and the observer, it was only natural that Everett would use Shannon's 
information theory to characterise quantum measurements within his relative state interpretation 
of quantum physics~\cite{Everett}. The novelty of Everett's approach was that it became possible to discuss the information in quantum measurements without recourse the Born projection postulate.
In this view, 
a quantum measurement is nothing but the establishment of correlations between the system and the 
measuring device. Everett did not discriminate between quantum and classical correlations, though he 
implicitly worked only with classical correlations as we now understand them. 

\section{Everett's View of Correlations}
\label{sec:eve}

According to the relative state view of quantum physics, a quantum measurement never gives conclusive results, but instead establishes correlations between the system and the measuring device.
As an example, for a system of a single qubit a typical Everettian style measurement therefore takes initially uncorrelated states of the system and measuring device to the post-measurement, correlated state
\begin{align}
\ket{\psi_\textrm{AB}} = \alpha \ket{0}\ket{m_1}+ \beta \ket{1}\ket{m_2},
\label{eq:postmeas}
\end{align}
where $\ket{0}$ and $\ket{1}$ are orthogonal states of the system, $\ket{m_1}$ and $\ket{m_2}$ are states of the measuring device with overlap $\langle b_0|b_1\rangle = \epsilon$ and $|\alpha|^2+|\beta|^2=1$. Intuitively, the degree of correlation between the system and measurement device depends upon $\epsilon$. This intuition can be quantified within the framework of Shannon's information theory by first introducing
one observable pertaining to the system $\hat X=x_1\ket{x_1}\bra{x_1}+x_2\ket{x_2}\bra{x_2}$, and one observable pertaining to the measuring device $\hat Y=y_1\ket{y_1}\bra{y_1}+y_2\ket{y_2}\bra{y_2}$.
The outcomes of measurements of these observables give rise to the two probability distributions, $p(x_i) = |\langle \psi_\textrm{AB} |x_i\rangle|^2$ and $p(y_j) = |\langle \psi_{AB} |y_j\rangle|^2$ as well as the joint distribution $p(x_i,y_j) = |\langle \psi_\textrm{AB} \ket{x_i}\ket{ y_j}|^2$. With this we can straightforwardly quantify the uncertainty in a given observable by the Shannon entropy Eq.~\eqref{eq:firstshannonentropy} and the correlation between the observables $\hat X$ and $\hat Y$ via the Shannon mutual information Eq.~\eqref{eq:shanmut}.

At this point, we make the important remark that while Everett showed that Shannon's information theory is capable of quantifying some correlations in quantum theory, it does not tell us the whole story; While the Shannon mutual information can tell us about correlations between the observables $\hat X$ and $\hat Y$, the correlation in post measurement quantum state itself Eq.~\eqref{eq:postmeas}, for instance, must be quantified using the quantum generalisation of Shannon's information theory. This will be the focus of Secs.~\ref{sec:Lindblad}-\ref{sec:Discord}.

Everett went on to establish the properties that any must be possessed by any `good' measure of correlations between two random variables. Namely, he proved that if either or both of the variables
undergo a {\em local} stochastic evolution, then the amount of
correlations between them can not increase  (in fact it usually decreases, see also Ref.~\cite{Penrose}).
As far as quantum measurements are concerned, this describes the physically intuitive fact that correlations established during the 
quantum measurements cannot be increased, i.e., the measurement
cannot be made to yield more information, by further processing the system and the apparatus separately. 
We note however, that when local stochastic processes are
correlated they virtually become global, and therefore the
correlations between the systems can increase as well as decrease.
A generalisation of this logic is also central to the theory of measuring quantum correlations and the
paradigm of local operations and classical communication (LOCC). We will return to this in Sec.~\ref{sec:entang}. Next, however, we discuss another important step in the development of correlations in quantum physics, due to Lindblad.

\section{Lindblad's Analysis of Quantum Measurements}
\label{sec:Lindblad}

Lindblad built on Everett's initial work by generalising his theory of quantum measurements to encompass mixed states in contact with an environment~\cite{Lindblad}.
In doing so, Lindblad employed the quantum versions of many of quantities from Shannon's information theory.  The formulation of classical
probability theory that is most naturally generalized to quantum
states is provided by Kolmogorov~\cite{Kolmogorov}, and an exposition expressing similarities with von Neumann's Hilbert
Space formulation~\cite{vonN} due to Mackey
\cite{mackey} (see also Ref.~\cite{Holevo}). 
For our purposes, however, it suffices to introduce the relevant quantities {\it ad hoc} by straightforward analogy with their classical counterparts.

First, for a quantum system described by a density matrix ${\rho}$, we define its von Neumann entropy by
\begin{align}
S_\textrm{N}(\rho) = -\mbox{Tr}\left[ \rho \ln \rho \right],
\label{eq:vonneuentropy}
\end{align}
This can be considered as the proper quantum
analogue of the Shannon entropy Eq.~\eqref{eq:firstshannonentropy} \cite{petz}.
For an observable $\hat{A}$ of the system described by $\rho$, the Shannon entropy $S(A)$, describing the uncertainties in the values of the observables,
is equal to the von Neumann
entropy $S_\textrm{N}(\rho)$ only when $[\hat A, \rho]=0$ i.e., $\hat A$ is a Schmidt observable. Otherwise,
\begin{align}
S(A) \ge S_\textrm{N}(\rho).
\nonumber
\end{align}
The physical interpretation of this result is that
there is more uncertainty in a single observable than in the whole of the
state, a fact which entirely contradicts our expectation.

The concept of mutual information Eq.~\eqref{eq:shanmut} can also be easily extended to quantum systems \cite{Vedral_RMP}.
For a bipartite state $\rho_{AB}$ and corresponding reduced states $\rho_A=Tr_B\left[\rho_{AB} \right]$ and $\rho_B=Tr_A\left[\rho_{AB} \right]$, the von Neumann mutual information between
the two subsystems is
defined as
\begin{align}
I_\textrm{N}(\rho_A:\rho_B ;\rho_{AB}) = S_\textrm{N}(\rho_A) + S_\textrm{N}(\rho_B) -
S_\textrm{N}(\rho_{AB}),
\label{def7b}
\end{align}
which refers to the correlation between the
whole subsystems rather than relating two observables only.
Following the Shannon mutual information, this quantity can
be interpreted as a distance between two quantum states by introducing the von Neumann relative entropy,  (in fact, this
quantity was first considered by Umegaki \cite{Umegaki}, but for
consistency reasons we name it after von Neumann).
The von Neumann relative entropy between two quantum states $\sigma$ and $\rho$ is defined as
\begin{align}
S_\textrm{N}(\sigma ||\rho) = \mbox{Tr} \left[ \sigma (\ln \sigma - \ln \rho)\right] .
\label{def8}
\end{align}
Now, the von Neumann mutual information Eq.~\eqref{def7b} can be understood as a distance of
the state $\rho_{AB}$ to the uncorrelated state $\rho_A\otimes\rho_B$,
\begin{align}
I_\textrm{N}(\rho_A:\rho_B ;\rho_{AB}) = S_\textrm{N}(\rho_{AB} || \rho_A\otimes\rho_B). \label{eq:vnmutual}
\end{align}
Lindblad's main contribution was to calculate and compare the classical and quantum mutual information during a measurement. This analysis was motivated by thermodynamical considerations. A question he considered is
how much entropy is generated during a quantum measurement. The entropy increase is fundamentally 
linked with the irreversibility of the quantum measurement and the correlations created between the system
and an apparatus. We will return to the connection between correlations and thermodynamics in Sec.~\ref{sec:harmonicq}.   

We now discuss a relation concerning the entropies of two
subsystems. One part of it is somewhat analogous to its classical
counterpart, but instead of referring to observables it is related
to the two states. The relation is the Araki-Lieb
inequality~\cite{Araki}. 
\begin{align}
S_\textrm{N}(\rho_A) + S_\textrm{N}(\rho_B) \ge S_\textrm{N}(\rho) \ge |S_\textrm{N}(\rho_A) - S_\textrm{N}(\rho_B)|,
\label{eq:lieb}
\end{align}
which can be considered one of the most important
results in the quantum theory of correlations.
Physically, Eq.~\eqref{eq:lieb} implies that their is more
information (less uncertainty) in an entangled state than if the
two states are treated separately. One the one hand, this arises naturally since by
treating the subsystems separately we have neglected the
correlations between them. However, 
to appreciate the extent to which this is a
counter-intuitive result we consider the following example.
Suppose a two level atom is interacting with a single mode of an
EM field as in the Jaynes-Cummings model~\cite{jc}. If the overall state is initially pure, and the whole
system is isolated then the entropies of the atom and the field
are equally uncertain at all the times. But this is not expected
since the atom has only two degrees of freedom and the field
infinitely many! This, is possible however, as by the second
observation, two level atom, is only
entangled with two dimensions of the field.

\section{Quantifying Entanglement}
\label{sec:entang}

The global state of a composite system is entangled if it may not be written as a product of the individual subsystems \cite{Werner89,Horodecki}. This implies that though the global state of the composite system may be known with certainty,
the state of each individual subsystem is uncertain, or, as Schr\"{o}dinger put it, ``The best possible knowledge of a whole does not include the best possible knowledge of its parts" \cite{Schrodinger} (c.f. Eq.~\eqref{eq:lieb}).
This reasoning can be quantified by considering the two qubit state
\begin{align}
\ket{\Psi_{AB}}= \cos\frac{\theta}{2}|0_\textrm{A}0_\textrm{B}\rangle +\sin\frac{\theta}{2}|1_A1_\textrm{B}\rangle,
\label{eq:entangledstate}
\end{align}
which, following the above definition is entangled for $0 < \theta < \pi$. For pure bipartite states, the amount of entanglement present is uniquely quantified by the von Neumann entropy Eq.~\eqref{eq:vonneuentropy} of (either of) the subsystems~\cite{Vedral1}.
Though the von Neumann entropy of the global state is zero, that of each of the reduced states is finite and determined by $\theta$,
\begin{align}
S_\textrm{N}(\rho_\textrm{A})=S_\textrm{N}(\rho_\textrm{B})=-\cos^2\frac{\theta}{2}\ln \cos^2\frac{\theta}{2}-\sin^2\frac{\theta}{2}\ln \sin^2\frac{\theta}{2}.
\end{align}
The entanglement is thus the lack of information concerning the reduced states of the subsystems in light of maximal information regarding the global state. Conversely, the more entangled the global state is the less information we have concerning the states of the subsystems.

For a seperable state ($\theta=0, \pi$), the von Neumann entropy of the global and reduced states are identically zero.
Conversely, when $\cos(\theta/2)=\sin(\theta/2)=1/\sqrt{2}$, the state Eq.~\eqref{eq:entangledstate} is maximally entangled and its reduced states are maximally mixed i.e., $S_\textrm{N}(\rho_\textrm{A})=S_\textrm{N}(\rho_\textrm{B})=\ln 2$. It  follows that the von Neumann mutual information Eq.~\eqref{def7b} of a maximally entangled state $I_\textrm{N}(\rho_\textrm{A};\rho_\textrm{B}:\rho_\textrm{AB}) = 2ln2$, is twice the maximum possible value of the Shannon mutual information between two random variables. Lindblad was the first to attribute this stronger-than-classical correlation to entanglement, though this assertion is only accurate for pure bipartite states and in general, entanglement does not encapsulate all forms of non-classical correlation. Note that for other values of $0<\theta<\pi$, the state is entangled but the von Neumann mutual information may not exceed the maximum possible Shanon mutual information.

More generally, for finite dimensional N-partite states, entanglement presents any form of correlation that cannot be captured by a separable state of the form
\begin{align}
\sigma=\sum_i p(i) \pi^{(i)}_1 \otimes \hdots \otimes \pi^{(i)}_\textrm{N},
\end{align}
where $\pi_n$ is the reduced density matrix of the $N^\textrm{th}$ subsystem. Operationally speaking, seperable states are those that are preparable under the paradigm of LOCC, while all other states are entangled \cite{Popescu, Bennett}. The entanglement in a quantum state can thus be quantified by its distance to the closest separable state \cite{Vedral1,Vedral2}. This distance can be expressed in a number of ways but the related details will not trouble us at present (see e.g. Ref. \cite{Amico}).
For infinite dimensional (continuous variable) systems, the theory of entanglement for the physically relevant set of Gaussian states is also well developed \cite{Kimmono, PlenioEisert}.

Entanglement aside, there are other forms of correlation in quantum systems with no counterpart in classical information theory. 
Just as with entanglement, the degree of other quantum correlations in a given state may be measured by its closest distance to other sets of states, which we briefly summarize here~\cite{Modi}. First, the set of classical states have the general form
\begin{align}
\chi=\sum_{k_1, \hdots ,k_\textrm{N}} p({k_1, \hdots ,k_\textrm{N}}) \ket{k_1\hdots k_\textrm{N}}\bra{k_1 \hdots k_\textrm{N}}.
\label{eq:classcorrstate}
\end{align}
This simply means that for a bipartite state of subsystem $A$ spanned by the orthogonal  basis $\{ |a_i\rangle \}$, and a subsystem $B$ spanned by the orthogonal basis $\{ |b_j\rangle \}$, the state $\chi=\sum_{i,j}p({a_i,b_j})\ket{a_i}\bra{a_i}\otimes \ket{b_j}\bra{b_j}$ is classically correlated when one or more of the associated joint probabilities do not factorise as $p(a_i,b_j)=p(a_i)\times p(b_j)$.

The states containing zero correlations, either quantum or classical, are the product states of the form
\begin{align}
\pi=   \pi_1\otimes \hdots \otimes \pi_\textrm{N}.
\end{align}
Note that the product states are a subset of the classical states which in turn are a subset of the separable states. The need to discriminate between separable states and classically correlated states follows from the 
existence of more forms of correlations than just entanglement and classical correlations.
This point is discussed further in the following section.

\section{Quantifying Discord}
\label{sec:Discord}

Quantum discord describes all forms of correlations beyond those encapsulated by classically correlated states. 
For a system of two qubits, $A$ and $B$, the amount discord they share can be quantified using an entropic measure by an extension of Shannon's original logic:
The more correlated  $A$ and $B$ are, the more we can learn about one by measuring the other. Accordingly, the degree of correlation between $A$ and $B$ is quantified by the reduction of entropy of $A$ ($B$) when $B$ ($A$) is measured.

Measurement of the state of subsystem $A$ is described by a positive operator value measure (POVM) 
whose elements $E_j$ satisfy the completeness relation $\sum_j E_j=1$. For each 
measurement outcome $j$, occurring with probability $p_j=\textrm{tr}\left(E_j \rho_{AB}\right)$, the 
post-measurement state of $B$ is $\rho_B^{(j)}=\textrm{tr}_A\left(E_j \rho_{AB}\right)/p_j$. The total 
classical correlations in the state $\rho_{AB}$ are then simply the maximum over all possible 
measurements performed on $A$ of the quantity~\cite{Henderson, Vedral-PRL}
\begin{align}
C(\rho_{AB}) = S_\textrm{N}(\rho_B) - \sum_j p_j S_\textrm{N}(\rho_B^{(j)}).
\label{eq:classcorr}
\end{align}
This quantity can be considered as one possible quantum generalisation of the classical mutual information Eq.~\ref{clascap}.
The classical information can also be defined by swapping the roles of $A$ and $B$, but the subtleties related to the question of symmetry of correlations in this context will not be relevant for our present discussion.

The quantum discord of $\rho_{AB}$ is then defined as the difference between the total correlation in the state, given by the von Nemann mutual information Eq.~\eqref{eq:vnmutual}, and the classical correlations \cite{Olliver, Zurek}
\begin{align}
\mathcal{D}(B|A)&=I(A:B)-\max_{\{E_i\}} \left[C(\rho_{AB})\right]
\nonumber \\
&=\min_{\{E_i\}}\left[I(A:B)- C(\rho_{AB})\right],
\label{eq:discord}
\end{align}
where, as previously mentioned, the maximization is performed over all possible POVMs acting on $A$.

The discord is thus the difference between two forms of the mutual information that are equivalent in classical information theory. Hence for classically correlated states the discord is zero, as expected. The discrepancy between these two quantities in quantum information theory arises from the role of measurement. While, classically, measurement serves to reveal the objective property of the system, in quantum mechanics the outcomes of measurement are basis dependent and affect the state of the system.

More generally, following the discussion of Sec.~\ref{sec:entang}, for a general N-partite state the discord is quantified by its distance to the closest classically correlated state Eq.~\eqref{eq:classcorrstate}. A theory of discord in continuous variable Gaussian states has also been developed~\cite{adesso}.
Importantly,  both entangled and seperable states can have finite discord. Consequently, the preparation of states with finite discord is possible under LOCC.
This relative ease of preparation has lead to questions regarding the technological implications of more general forms of quantum correlations than just entanglement. Beyond this, the recent level of interest in discord has taught us to discriminate between different forms of correlations, a distinction that has no counterpart in classical information. Discord has also received much attention in attempts to quantify the ``quantumness'' of a physical system or process. In the next section we give one such application in the context of quantum non-equilibrium thermodynamics.

\section{Correlations in quantum thermodynamics} 
\label{sec:harmonicq}

We consider the generation of correlations in two coupled thermal harmonic
oscillators subject to a sudden change of their harmonic coupling.
Starting with a classical Hamiltonian description, we derive the average irreversible work done during the change by treating the time dependence of the coupling as a thermodynamic transformation.
Next, the quantum irreversible work is calculated from an analogous quantised Hamiltonian and transformation.
In both cases, we assume there is initially zero harmonic coupling and each oscillator is prepared in a Gibbs state, there hence being zero correlations between the oscillators. The transformation of the classical system leads to the generation of classical correlations only, while, in the quantum case, both classical and non-classical correlations are established between the oscillators. We investigate if these more general forms of correlations have
a straightforward thermodynamic meaning by investigating how the Gaussian discord scales with the difference between the quantum and classical irreversible work. For a study of the entanglement properties of coupled thermal oscillators see Refs.\cite{Galve, Plenio}.

To begin, we consider two classical harmonic oscillators of identical mass $m$ and natural
frequency $\omega$ with time-dependent coupling $\lambda(t)$,
\be
H(\lambda)= \frac{1}{2m}(p^2_{1}+p^2_{2})+\frac{m \omega^2}{2}
(q_1^2+q_2^2)+\frac{m \lambda^2(t)}{2}(q_1-q_{2})^2,
\label{eq:classicalhamiltonian}
\ee
where $q$ and $p$ denote position and momentum respectively.
Initially, the system is allowed to thermalise with a heat bath at
temperature $\beta \equiv 1/k_B T$, where $k_B$ is Boltzmann's constant. The partition function of the system is defined as \cite{Bloch},
\begin{align}
Z_C(\lambda)&:=\frac{1}{h^2}
\int\mathrm{d}{\bf q}\mathrm{d}{\bf p}\e{-\beta H(\lambda)}
\nonumber \\
&=\left(\frac{2\pi}{\beta\omega}\right)^2\frac{1}{ \sqrt{1+2\lambda^2/\omega^2}}.
\nonumber
\end{align}
where ${\bf q} \equiv (q_1,q_2), {\bf p} \equiv (p_1,p_2)$ and $h$ has
dimensions of Planck's constant, making the partition function dimensionless.

The system is prepared by turning off the harmonic coupling ($\lambda=0$) and allowing the
oscillators to re-thermalise before the contact with the heat bath is removed. At $t=0$ the system is therefore described by the Gibbs distribution,
\be
\wp({\bf q},{\bf p})=\frac{\e{-\beta H(0)}}{Z_C(0)},
\label{eq:classgibbs}
\ee
where $Z_C(0) =(2 \pi  / h \beta \omega_0)^2$ is the partition function
in the presence of zero harmonic coupling.

Next, the harmonic coupling is instantaneously switched to the new value $\lambda=\lambda_0$. In the presence of zero heat flow, the first law of thermodynamics asserts that the work done on the system under
this protocol is simply the change of internal energy
\begin{align}
W_C&=H(\lambda_0)-H(0)
\nonumber \\
&=\frac{m\lambda^2_0}{2}(q_1-q_{2})^2.
\nonumber
\end{align}
The average work, over many realisations of this protocol, is obtained by
averaging $W_C$ over the initial Gibbs distribution Eq.~\eqref{eq:classgibbs}, thus
\begin{align}
\langle W_C \rangle&=\int\mathrm{d}{\bf q}\mathrm{d}{\bf p} \; \wp({\bf q},{\bf
p}) W_C
\nonumber \\
&=\frac{1}{Z_C(0)}\int\mathrm{d}{\bf q}\mathrm{d}{\bf p}
\;\e{-\beta\left(\frac{1}{2m}(p^2_{1}+p^2_{2})+\frac{m\omega^2}{2}(
q_1^2+q_2^2)\right)} \frac{m\lambda^2_0}{2}(q_1-q_{2})^2
\nonumber \\
&=\frac{ \lambda_0^2}{\beta \omega^2}.
\label{eq:classwork}
\end{align}
With a little extra effort we are able to calculate the average irreversible (or
`dissipated') work,
\begin{align}
\langle W_C^\textrm{irr} \rangle=\langle W_C \rangle -\Delta F_C
\label{eq:classirrwork}
\end{align}
where the free energy change of the process is defined as
\begin{align}
\Delta F_C&= -\frac{1}{\beta}\ln \frac{Z_C(\lambda_0)}{Z_C(0)}
\nonumber \\
&=\frac{1}{2\beta}\ln\left[1+\frac{2\lambda_0^2}{\omega^2}\right],
\nonumber
\end{align}
completing the analysis of the classical system.

The treatment of the analogous quantum system begins with canonical quantisation of the classical Hamiltonian Eq.~\eqref{eq:classicalhamiltonian}, thus
\begin{align}
\hat{H}(\lambda)=
\frac{1}{2m}(\hat{p}^2_{1}+\hat{p}^2_{2})+\frac{m\omega^2}{2}(
\hat{q}_1^2+\hat{q}_2^2)+\frac{m\lambda^2(t)}{2}(\hat{q}_1-\hat{q}_{2})^2,
\label{eq:quanthamil}
\end{align}
where, now, the position and momentum operators obey the canonical  commutation relation $\left[\hat{x}_j,\hat{p}_k\right]=\delta_{j,k}i\hbar$, and $\hbar$ is the reduced Planck's constant. The quantum Hamiltonian Eq.~\eqref{eq:quanthamil} is diagonalised by application of the unitary operator (see Ref.~\cite{kim} for details)
\begin{align}
\hat{T}=\textrm{exp}\left(\frac{i\pi}{4\hbar}\left(\hat{q}_1\hat{p}_2-\hat{p}_1\hat{q}_2 \right)\right).
\end{align}
Imagining for a moment that $\hat{x}_{1,2}$ and $\hat{p}_{1,2}$ represent the quadratures of two modes of the electromagnetic field, the operator $\hat{T}$ has the physical interpretation of mixing the two modes at a 50:50 beam-splitter.
Under this operation, the Hamiltonian Eq.~\eqref{eq:quanthamil} becomes
\begin{align}
\hat{H}'=\hat{T}\hat{H}\hat{T}^\dagger =\frac{1}{2m} \hat{p}_1^2+\frac{m\omega_1^2}{2}\hat{q}_1^2+
\frac{1}{2m}\hat{p}_1^2 +\frac{m\omega_2^2}{2}\hat{q}_2^2,
\label{eq:diagham}
\end{align}
where the harmonic coupling is encoded in the renormalized frequencies $\omega_1(\lambda)=\omega_0 \sqrt{1-\lambda^2/\omega_0^2}$, $\omega_2(\lambda)=\omega_0 \sqrt{1+\lambda^2/\omega_0^2}$ and $\omega_0=\sqrt{\omega^2+\lambda^2}$, thus $\omega_1(0)=\omega_2(0)=\omega$. 

Finally, defining the creation and annihilation operators
\begin{align}
\hat{a}_k(\lambda)&:=\sqrt{\frac{m\omega_k}{2\hbar}}\left(\hat{q}_k+\frac{i}{m\omega_k}\hat{p}_k \right),
\nonumber \\
\hat{a}_k^\dagger(\lambda)&:=\sqrt{\frac{m \omega_k}{2\hbar}}\left(\hat{q}_k-\frac{i}{m\omega_k}\hat{p}_k \right),
\end{align}
that obey the commutation relation $\left[\hat{a}_j,\hat{a}_k^\dagger \right]=\delta_{j,k}$ $\forall \lambda$, Eq.~\eqref{eq:diagham} can be rewritten as
\begin{align}
\hat{H}'=\hbar \omega_1\left(\hat{a}^\dag_1\hat{a}_1+ \frac{1}{2}\right)+\hbar \omega_2 \left(\hat{a}^\dag_2\hat{a}_2 +\frac{1}{2}\right).
\label{eq:creationahm}
\end{align}
Eqs.~\eqref{eq:diagham} and \eqref{eq:creationahm} have the familiar form of two uncoupled harmonic oscillators with natural frequencies $\omega_1$ and $\omega_2$ respectively.
Consequently the partition function of the system at inverse temperature $\beta$ is simply the product of the partition functions for each individual oscillator i.e.,
\begin{align}
Z_Q(\lambda):=\tr{\e{-\beta \hat{H}}}=\tr{\e{-\beta \hat{H}'}}=\frac{1}{4}\textrm{csch}\left( \frac{\beta \hbar  \omega_1}{2}\right)\textrm{csch} \left( \frac{\beta \hbar \omega_2}{2} \right).
\nonumber
\end{align}
where $\textrm{tr}[.]$ denotes the trace over the two-mode Fock basis spanning the Hilbert space of the Hamiltonian.

To proceed,
we consider an identical protocol to that performed upon the classical system:
Initially, the harmonic coupling is zero ($\lambda=0$) and the system is prepared in the
corresponding Gibbs state at inverse temperature $\beta$,
\be
\rho_\textrm{G}=\frac{1}{Z_Q(0)}\e{-\beta \hat{H}(0)},
\label{eq:initialqstate}
\ee
where the appropriate partition function is explicitly $Z_Q(0)=(1/4)\textrm{csch}^2(\beta\hbar \omega/2)$.
Subjecting the oscillators to an identical thermodynamic transformation, taking the harmonic
coupling to the final value $\lambda=\lambda_0$,
we define the following object,
\begin{align}
\hat{W}_Q&=\hat{H}(\lambda_0)-\hat{H}(0)
\nonumber \\
&=\frac{m\lambda_0^2}{2}(\hat{q}_1-\hat{q}_{2})^2
\nonumber \\
&=\frac{\hbar\lambda_0^2}{2\omega}(\hat{a}_1^\dagger(0)+\hat{a}_1^\dagger(0)-\hat{a}_2(0)-\hat{a}_2^\dagger(0))^2.
\label{eq:workop}
\end{align}
Note that, in general thermodynamic transformations of quantum systems, it is
not possible to define a so-called `operator of work' \cite{Hanggi}. However for a sudden
quench of the harmonic coupling, Eq.~\eqref{eq:workop} allows
definition of the average work over many realisations of the protocol as,
\begin{align}
\langle \hat{W}_Q \rangle &= \tr{\rho_\textrm{G} \hat{W}_Q},
\nonumber \\
&=\frac{1}{Z_Q(0)}\frac{\hbar\lambda^2_0}{2\omega}\tr{\e{-\beta \hbar \omega
\left(\hat{a}_1^\dagger(0) \hat{a}_1(0)+\hat{a}_2^\dagger \hat{a}_2+1\right)}\left(\hat{a}_1^\dagger(0)
\hat{a}_1(0)+\hat{a}_2^\dagger \hat{a}_2+1 \right)},
\nonumber \\
&=\frac{\hbar \lambda^2_0}{2 \omega}\coth\left(\frac{\beta
\hbar \omega}{2} \right)
\label{eq:quantwork}
\end{align}
Again, we are able to calculate the average irreversible work,
\be
\langle W_Q^\textrm{irr} \rangle =\langle \hat{W}_Q\rangle-\Delta F_Q,
\label{eq:quantirrwork}
\ee
where the free energy change is,
\begin{align}
\Delta F_Q&=-\frac{1}{\beta}\ln \frac{Z_Q(\lambda_0)}{Z_Q(0)}
\nonumber\\
&= -\frac{1}{\beta}\ln\frac{\textrm{csch}\left( \beta \hbar \omega_1(\lambda_0)/2\right)\textrm{csch}\left( \beta \hbar \omega_2(\lambda_0)/2\right)}{\textrm{csch}^2 \left( \beta \hbar \omega/2 \right)}.
\label{eq:quantfree}
\end{align}
The thermodynamic significance of the the quantities Eqs.~\eqref{eq:quantwork} - \eqref{eq:quantfree} can be understood as follows:
For a quasistatic transformation, the irreversible work is zero and the total work is given simply by the free energy change. The free energy change is thus the energy required to transform between spectra of the initial and final Hamiltonian. For non-quasistatic processes, however, the average irreversible work is finite and gives the energy expended in exciting the system during the transformation.

In the quantum regime $k_B T<<\hbar \omega_0$ the expressions for the quantum and classical irreversible work (Eqs.~\eqref{eq:quantirrwork} and \eqref{eq:classirrwork}, respectively) differ. We thus define the average excess quantum dissipated work as
\begin{align}
\Omega&=\langle W^\textrm{diss}_Q \rangle -\langle W^\textrm{diss}_C \rangle,
\nonumber \\
&=\frac{\hbar \lambda^2}{2 \omega}\left(\coth\frac{\beta \hbar \omega}{2}-\frac{2}{\beta \hbar \omega}\right)+\frac{1}{\beta}\ln\frac{\textrm{csch}\left( \beta \hbar \omega_1(\lambda_0)/2\right)\textrm{csch}\left( \beta \hbar \omega_2(\lambda_0)/2\right)}{\textrm{csch}^2 \left( \beta \hbar \omega/2 \right)}+\frac{1}{2\beta}\ln\left[1+\frac{2\lambda_0^2}{\omega^2}\right].
\nonumber
\end{align}
Clearly, $\Omega \geq 0$, with equality reached in the classical limit $k_B T> >\hbar \omega_0$ or for a quasistatic process. The average excess quantum dissipated work is thus the additional energy expended in exciting a quantum system relative to its classical analogue undergoing an identical transformation.

We investigate whether this quantity also encodes the fact that during the transformation, the two components of the quantum system share additional, more general forms of correlations than available to its classical analogue. As a starting point, following Ref.~\cite{adesso} we calculate the Gaussian discord induced by the sudden switch on of the harmonic coupling. Explicitly, we consider the discord $\mathcal{D}$ generated in transforming an initial product of identical thermal states Eq.~\eqref{eq:initialqstate} according to the unitary evolution $U=\e{i\hat{H(\lambda_0)}}$ generated by the Hamiltonian Eq.~\eqref{eq:quanthamil} i.e., $\mathcal{D}(U \rho_G U^\dagger)$. Note that as the Hamiltonian is bilinear in the creation and annihilation operators, it preserves the Gaussian character of the initial thermal states.

\begin{figure}
\includegraphics[width=0.75\linewidth]{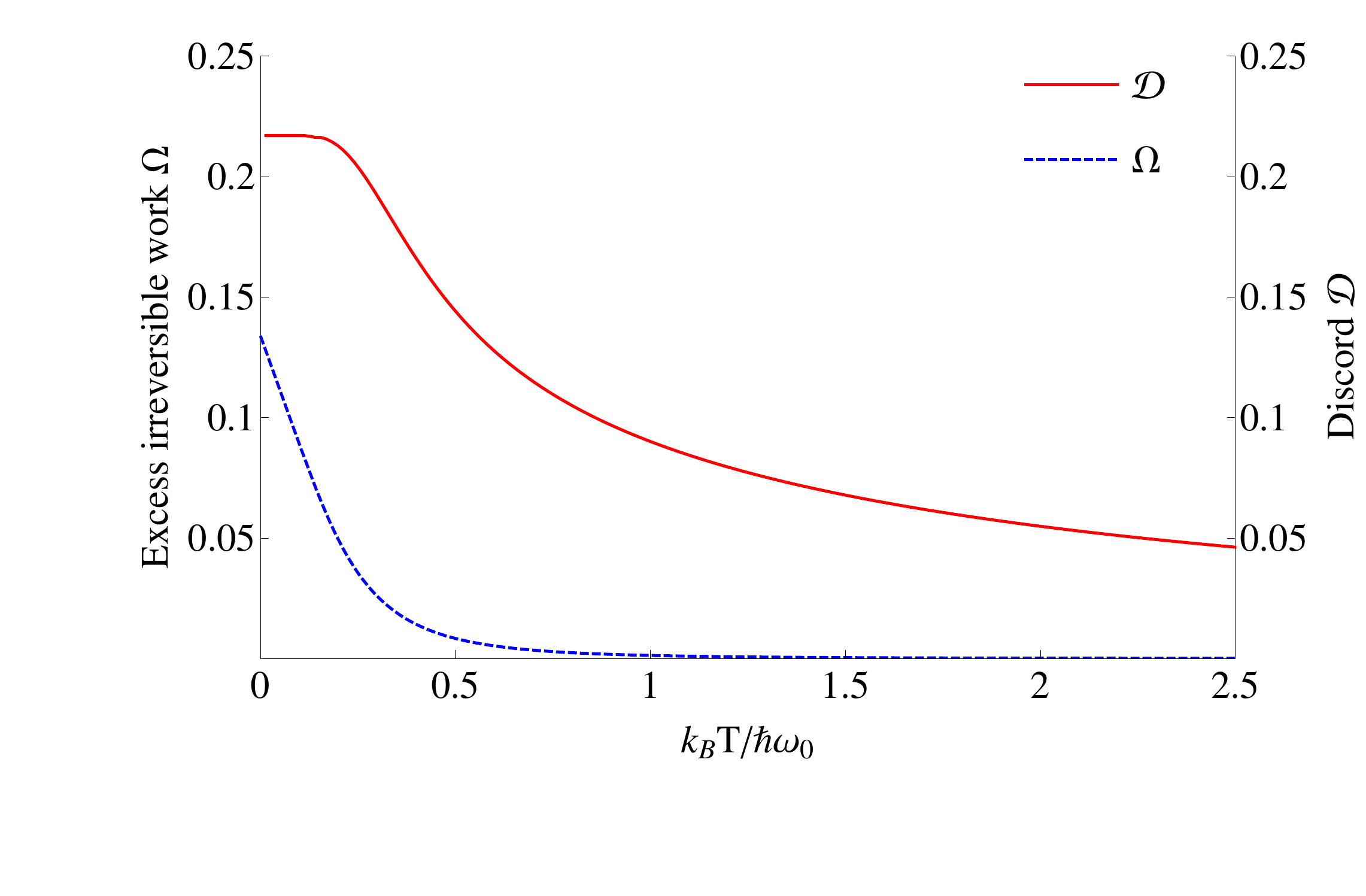}
\caption{The average excess quantum dissipated work (dotted line) and quantum discord (solid line) following a sudden quench of the harmonic coupling with amplitude $\lambda_0=\omega$ and $\hbar=k_B=\omega=m=1$, as a function of temperature.}
\label{fig:figure1}
\end{figure}

At this stage, without loss of generality we set $k_B=\hbar=\omega=m=1$, and plot (Fig.~\ref{fig:figure1}) the average excess quantum work and Gaussian discord for a quench amplitude $\lambda_0=\omega$ as a function of temperature $T$.
Fig.~\ref{fig:figure1} shows that the discord and excess quantum irrevesible work
 show qualitatively similar behaviour, decaying with increasing temperature as expected. However, it is not clear if there exists a definitive link between the two quantities. While the excess quantum irreversible work should contain some information on the thermodynamic cost of correlation generation, this is likely to be convoluted with contributions to the total from other sources, e.g. exciting the oscillators individually without correlating them. Nonetheless, it remains an interesting open question as to what the relation of quantum discord, and other forms of correlation, to non-equilibrium thermodynamics may be.

%

\section*{Acknowledgments}
The authors thank John Goold, Janet Anders, Kavan Modi, Myungshik Kim and Marco Genoni for interesting and helpful
discussions relating to the topic of this work. RD is funded by the EPSRC. VV is a fellow of Wolfson College Oxford and is supported by the John Templeton Foundation, the National Research Foundation, the Ministry of Education in Singapore and the Leverhulme Trust (UK).

\end{document}